# Proximity-induced superconductivity in nanowires: Mini-gap state and differential magnetoresistance oscillations

Jian Wang[*], Chuntai Shi, Mingliang Tian, Qi Zhang, Nitesh Kumar, J. K. Jain, T. E. Mallouk, M. H. W. Chan[*]

*The Center for Nanoscale Science and Department of Physics, The Pennsylvania State University, University Park, Pennsylvania 16802-6300, USA*

\* Corresponding authors: juw17@psu.edu (Wang); chan@phys.psu.edu (Chan).

We study proximity-induced superconductivity in gold nanowires as a function of the length of the nanowire, magnetic field, and excitation current. Short nanowires exhibit a sharp superconducting transition, whereas long nanowires show nonzero resistance. At intermediate lengths, however, we observe two sharp transitions; the normal and superconducting regions are separated by what we call the mini-gap phase. Additionally, we detect periodic oscillations in the differential magnetoresistance. We provide a theoretical model for the mini-gap phase as well as the periodic oscillations in terms of the coexistence of proximity-induced superconductivity with a normal region near the center of the wire, created either by temperature or application of a magnetic field.

If a superconductor (S) is brought into contact with a normal metal (N), the Cooper pairs do not disappear abruptly at the S-N interface but leak into the normal metal. This extends superconducting behavior into the normal metal and simultaneously weakens the superconductivity near the interface of the S-N structure [1-3]. Most of the recent studies on the "proximity effect" have been carried out in sub-micron and nanometer scale S-N structures [4-12]. In this paper, we report on our experimental studies of proximity-induced superconductivity in individual high quality single crystal Au nanowires, with special focus on the behavior as a function of the length of the nanowire. A sharp superconducting transition was observed in a 1 μm Au nanowire. The resistance of a 1.9 μm sample exhibits a gradual decay with decreasing temperature below an onset temperature but remains finite down to the lowest temperatures of our experiment. The most surprising features in our study are: (i) For the 1.2 μm nanowire the resistance drops from normal to zero in two distinct steps, indicating the existence of a new state with a small superconducting gap for nanowires of intermediate length. In the mini-gap region, two distinct transitions are also observed as a function of the magnetic field. (ii) We observe unusual periodic oscillations in the differential magnetoresistance in both 1 and 1.2 μm wires.

Crystalline Au nanowires with a diameter of 70 nm were fabricated by electrodepositing Au in the pores of track-etched polycarbonate membrane [13]. Free-standing nanowires were 'harvested' by dissolving the polycarbonate membrane in dichloromethane and were precipitated from the solvent in a centrifuge. The nanowires were then stored as a suspension in dichloromethane. A high-resolution transmission electron microscope (TEM) image shows the crystalline structure at the sub-nanometer scale (inset of Fig. 1(a)). To measure an individual Au nanowire, we placed a drop of the nanowire suspension on a silicon substrate with a 1 μm thick $Si_3N_4$ insulating layer. The sample was then transferred into a commercial focused ion beam (FIB) etching and deposition system (FIB/SEM FEI Quanta 200 3D) for the attachment of electrodes [14, 15]. As shown in the top left inset of Fig. 1(a), four FIB-assisted W electrodes were deposited to contact an individual Au nanowire for a standard four-probe measurement. In the process of W electrode preparation, $Ga^+$ imaging was suspended to reduce gallium contamination of the Au nanowire. During the deposition, the chamber pressure was ~ $10^{-5}$ Torr and the deposition FIB current was set to below 20 pA. Here, the FIB-deposited W strips used as the electrodes (about 250 nm wide and 100 nm thick) were amorphous and composed of tungsten, carbon and gallium [16, 17]. A superconducting transition of 5.1 K was observed in the W strip by a standard four-probe measurement, as found in prior studies [16]. TEM showed that the diffusion of the W atoms along the Au wires is limited to within 200 nm from the edge of the W strips. In another control experiment, the resistance between two parallel W strips separated by less than 500 nm on a $Si_3N_4$ substrate was larger than $10^5$ Ω at 1.8 K, indicating an absence of direct transport between the W strips.

It should be noted that the gallium ions used in the FIB process are not responsible for the observed superconductivity since the FIB deposited Pt strip contains gallium but is not superconducting [14, 15].

Figure 1(a) shows resistance as a function of temperature (R-T) for individual 70 nm diameter crystalline Au nanowires of lengths 1 µm, 1.2 µm and 1.9 µm, defined as the distance $L$ between the inner edges of the two voltage electrodes. The vertical scale is normalized to the resistance at T=6 K which are, respectively, 196 Ω, 152 Ω and 100 Ω for the 1, 1.2 and 1.9 µm wires. The variation in the normal state resistance reflects the difference in the density of defects of the wires originating either during chemical deposition or the FIB processes. The excitation current in the measurement is 50 nA. All three wires show an onset of superconductivity near 4.5 K, slightly lower than that of the W electrode strip, which is expected since the proximity effect weakens the superconductivity near the S-N interface. Zero resistance, defined as a resistance smaller than the instrumental resolution of +/- 0.2 Ω, is found below T=4.05 K for the 1 µm sample. For the 1.9 µm sample, the resistance loses 40% of its normal state value by 4.22 K, but thereafter decays gradually and remains finite down to the lowest temperature (1.8 K) measured, extrapolating to 25% of the normal state value at T=0 K.

The resistance drop occurs in two steps for the 1.2 µm wire. In the first step between 4.5 K and 4.14 K, the resistance reduces sharply to 16% of its normal state value, followed by a much slower drop until 3.43K where the resistance vanishes. The two step drop in resistance is suggestive of two distinct transitions. Figure 1(b) shows the resistance as a function of temperature for the 1.2 µm Au nanowire at several magnetic fields. (The magnetic field was applied perpendicular to the axis of the nanowire in all data presented in this paper.) The two transitions move to lower temperatures with increasing field and merge into one at 30 kOe. Superconductivity in the nanowire is fully suppressed at 70 kOe.

The resistance vs magnetic field (R-H) plots for the three samples at several temperatures are shown in Fig. 2. The evolution of the R-H curves of the 1.0 µm sample (panel a) exhibits a well defined superconducting transition. The R vs H curve at 2 K for the 1.2 µm sample shown in panel b is similar to that for the 1 µm sample, but the scans at 3 K and 3.5 K show fascinating new features. Specifically, an additional mini-resistance valley is found in a narrow low magnetic field region. A magnified view of the evolution of this mini-resistance valley in the 1.2 µm wire as a function of temperature is shown in Fig. 2(d). At 2.4 K, there is no sign of the mini-resistance valley. At 2.5 K, the resistance suddenly jumps and then quickly drops back to zero at 2.5 kOe. At 2.6 K, this magnetic field symmetric resistance fluctuation becomes more clearly developed and the baseline of the fluctuating resistance at fields above the mini-valley region is seen to be smoothly increasing with field. Between 2.7 and 3.0 K, the resistance fluctuations evolve to the low field mini-resistance valley. Between 3.4 and 3.5 K, the zero field resistance changes from zero to a finite value with remnants of fluctuations. With increasing temperature the mini-resistance valley continues to shrink both in width and depth, disappearing at T ≥ 4 K; the primary resistance valley defining the critical field also decreases in width (see Fig. 2(b)). The appearance and the disappearance of this mini-valley feature is one of the principal findings of this paper. We note that a magnetic-field symmetric mini-resistance valley is also observed in the 1.9 μm sample (panel c), shrinking with increasing temperature; Fig. 2(e) shows the details of the mini-valley at 1.8 K.

Upon closer inspection, the magnetoresistance of the 1.0 and 1.2 μm wires shows, at low temperatures, small terraces. These are better revealed in the dR/d|H|-H plots (see Fig. 3). For both wires, dR/d|H| is found to be a smooth function of H at high temperatures. At T=4 K, the dR/d|H|-H curves show a field-symmetric broad peak in both wires (with different finer features). The broad peaks evolve outward with decreasing temperature, and exhibit periodic oscillations at low T. For both wires, the periodic nature of these oscillations is well developed between 2-3 K with roughly the same period, $\Delta H \sim 2.5 kOe$. At T=3.5 K, the oscillations remain but the periodic character is less pronounced. The two small peaks around zero field in the 1.2 μm wire in Fig. 3 (a) mark the boundaries of the mini-resistance valley. We also find oscillations in the 1.9 μm wire, but without any well defined periodicity. To ascertain that these oscillations originate from proximity-induced superconductivity in the Au nanowires, we performed control experiments on a W strip contacted with 4 W electrodes and did not see any oscillations, periodic or otherwise.

The phase boundary between the zero and finite resistance states in the $H-T$ plane for the 1.2 μm wire is shown in Fig. 4(a). The superconducting phase consists of two regions. Above T~2.5 K, the zero resistance state is destroyed by a very small magnetic field. At lower temperatures, the critical field for the wire is enhanced rapidly. Both phase boundaries are found to roughly follow the relation $H_C^2 \propto T_0^2 - T^2$. We also show $H_C^O$ and $H_C^S$ as a function of T in the same figure. $H_C^O$ is defined as the field where the resistance of the wire drops to 90 percent of its normal state value and $H_C^S$ is defined as the field where the H-dependence of the resistance switches from a slowly-varying function to a rapidly-increasing function (see Fig. 2(b)).

We next describe a qualitative theoretical picture that is consistent with many of the above observations. The proximity induced gap in the gold nanowire in zero magnetic field can be modeled as [1] (see Fig. 4(c))

$$\Delta(x) = \Delta_a \cosh(x/\xi_N)/\cosh(a/\xi_N), \qquad -a < x < a \qquad (1)$$

where $2a$ is the length of the gold nanowire, $\xi_N$ is its coherence length, and $\Delta_a$ is the gap at the boundary. We have two characteristic gaps in this model: a bigger gap $\Delta_a$ at the boundary and a smaller gap $\Delta_b = \Delta(x=0) = \Delta_a \cosh^{-1}(a/\xi_N)$ in the middle. It is natural to expect that when $\Delta_b$ is small compared to the temperature of the wire, the central part of the nanowire becomes normal. For very long nanowires, the middle part is always normal at the experimental temperatures, so no true superconductivity is seen. In very short nanowires, $\Delta_b$ is on the same order as $\Delta_a$, and superconductivity in the entire nanowire is destroyed simultaneously. However, for certain intermediate lengths, at the critical temperature (defined as the temperature where perfect conductivity is lost) only the middle region becomes normal, and then at a higher temperature, the whole nanowire becomes normal, thus producing two transitions. We believe that the 1.2 μm nanowire belongs to the last category.

Proximity effect induced superconductivity can be suppressed by a "breakdown magnetic field" $H_b$, which is much less than the critical field of the superconductor (W electrodes in this case). In the literature, an S-N bilayer film exposed to a magnetic field parallel to the interface has been studied [18], and we borrow those results for our system. The gap function for $H > H_b$ switches to a new solution:

$$\Delta(x) = \begin{cases} \Delta_a \dfrac{\sinh[(x-d/2)/\xi_N]}{\sinh[(a-d/2)/\xi_N]}, & d/2 < x < a \\ 0, & -d/2 < x < d/2 \\ \Delta_a \dfrac{\sinh[(d/2+x)/\xi_N]}{\sinh[(d/2-a)/\xi_N]}, & -a < x < -d/2 \end{cases} \quad (2)$$

where the length $d$ is a function of T and H (see Fig. 4(d)). This suggests the possibility of a first order transition at $H_b$, where the nanowire acquires a small but nonzero resistance, as observed for our 1.2 μm nanowire, which exhibits a sharp jump in resistance in the temperature range 2.6-3.4 K. At even higher applied fields, the superconductivity in the outer segments of the nanowire is suppressed, and the resistance increases gradually until it reaches the full normal value. In contrast, when $\Delta_b$ is relatively large, only one transition is expected as a function of the magnetic field, which we believe to be the case for the 1.0 μm nanowire, or for the 1.2 μm nanowire at very low temperatures. The model does not explain, however, the enhanced fluctuations in resistance for the 1.2 μm nanowire near $H \sim H_b$, or the magnetoresistance mini-valley in the 1.9 μm nanowire.

Increasing the current can also induce a transition from zero to a finite resistance state. One possible scenario is that $\Delta_b$ is destroyed at a critical current $I_C$ while $\Delta_a$ is intact, resulting in a relatively small resistance, while the whole nanowire is driven normal only at a larger current. The critical current is thus controlled by the smaller gap $\Delta_b$ via the relation

$$I_C \propto (\Delta^* \frac{d}{dx}\Delta - \Delta \frac{d}{dx}\Delta^*) \propto \Delta_b^2 \qquad (3)$$

For the 1.2 and 1.0 $\mu m$ nanowires, the critical current is found to behave (Fig. 4(b)) as $I_C \propto (T_C - T)^2$ with $T_C \sim$ 3.4 and 4.1 K, respectively. Again, for relatively large $\Delta_b$, there is a single transition to the normal state as a function of current. While our model is consistent with the qualitative behavior, we cannot explain the behavior $I_C \propto (T_C - T)^2$.

As for the periodic oscillations in the differential magnetoresistance, we appeal to the analogy of oscillations in the supercurrent in S-N-S junctions, which in the presence of a magnetic field is given by [19] $I_S \propto \sin(\Delta\phi - \frac{2e}{h}HS)$, where $HS$ is the magnetic flux through the wire, and $\Delta\phi$ is the difference in the phases of the superconducting order parameters on the two ends. In our model we also effectively have an S-N-S junction, where the N refers to the central part of the nanowire and S to the outer regions. In a simple picture, where the net current is a combination of normal and super currents, periodic oscillations are produced by the latter. With the experimental values of H~2.5 kOe and the diameter of the nanowires W=70 nm, we determine $d \approx$ 110nm, which is consistent with the expectation that $d$ should be only a small fraction of the length of the nanowire. However, the near independence of the period on H and T is inconsistent with the expectation that the length $d$ of the normal region depends on both these parameters.

Our results suggest that proximity-induced superconducting nanowires not only form a new platform for fundamental study of superconductivity in nanoscale systems but also offer a real opportunity for applications as superconducting interconnects in nanodevices. This work was supported by the Penn State MRSEC under NSF grant DMR-0820404 and the Pennsylvania State University Materials Research Institute Nano Fabrication Network and the National Science Foundation Cooperative Agreement No. 0335765, National Nanotechnology Infrastructure Network, with Cornell University.

**References**

[1] P. G. De Gennes, Rev. Mod. Phys. **36**, 225 (1964).
[2] B. Pannetier and H. J. Courtois, Low Temp. Phys. **118**, 599 (2000).
[3] F. S. Bergeret, A. F. Volkov, and K. B. Efetov, Rev. Mod. Phys. **77**, 1321(2005).
[4] S. Gueron, H.Pothier, Norman O. Birge, D. Esteve, and M. H. Devoret, Phys. Rev. Lett. **77**, 3025 (1996)**.**
[5] R. S. Decca, H. D. Drew, E. Osquiguil, B. Maiorov, and J. Guimpel, Phys. Rev. Lett. **85**, 3708 (2000).
[6] J. Haruyama, A. Tokita, N. Kobayashi, M. Nomura, S. Miyadai, K. Takazawa, A. Takeda, and Y. Kanda, Appl. Phys. Lett. **84**, 4714 (2004).
[7] M. Tian, N. Kumar, S. Xu, J. Wang, J. S. Kurtz, and M. H. W. Chan, Phys. Rev. Lett. **95**, 076802 (2005).
[8] Y. J. Doh, J. A. van Dam, A. L. Roest, E. P. A. M. Bakkers, L. P. Kouwenhoven, and S. De Franceschi, Science **309**, 272 (2005).
[9] J. Xiang, A. Vidan, M. Tinkham, R. M. Westervelt, and C. M. Lieber, Nature Nanotechnol. **1**, 208 (2006).
[10] H. B. Heersche, P. Jarillo-Herrerol, J. B. Oosinga, L. M. K. Vandersypen, and A. F. Morpurgo, Nature **446**, 56 (2007).
[11] H. D. Liu, Z. X. Ye, Z. P. Luo, K. D. D. Rathnayaka, and W. H. Wu, Physica C **468**, 304 (2008).
[12] H. le Sueur, P. Joyez, H. Pothier, C. Urbina, and D. Esteve, Phys. Rev. Lett. **100**, 197002 (2008)**.**
[13] M. Tian, J. Wang, J. Kurtz, T. E. Mallouk, and M. H. W. Chan, Nano Lett. **3**, 919 (2003).
[14] S. Valizadeh, M. Abid, F. Hernandez-Ramirez, A. R. Rodriguez, K. Hjort, and J. A. Shweitz, Nanotechnology **17**, 1134 (2006).
[15] N. Kumar, M. Tian, J. Wang, W. Watts, J. Kindt, T. E. Mallouk, and M. H. W. Chan, Nanotechnology **19**, 365704 (2008).
[16] E. S. Sadki, S. Ooi, and K. Hirata, Appl. Phys. Lett. **85**, 6206 (2004).
[17] W. X. Li, J. C. Fenton, Y. Q. Wang, D. M. McComb, and P. A. Warburton, J. Appl. Phys. **104**, 093913 (2008).
[18] Groupe de Supraconductivite d'Orsay, quantum Fluids, Proc. Of the Sussex Univ. Symp., 1965 (D.F. Brewer,de.), North-Holland, Amsterdam, 1966.
[19] A. A. Abriskosov, Fundamentals of the theory of metals (North-Holland, Amsterdam. 1988).

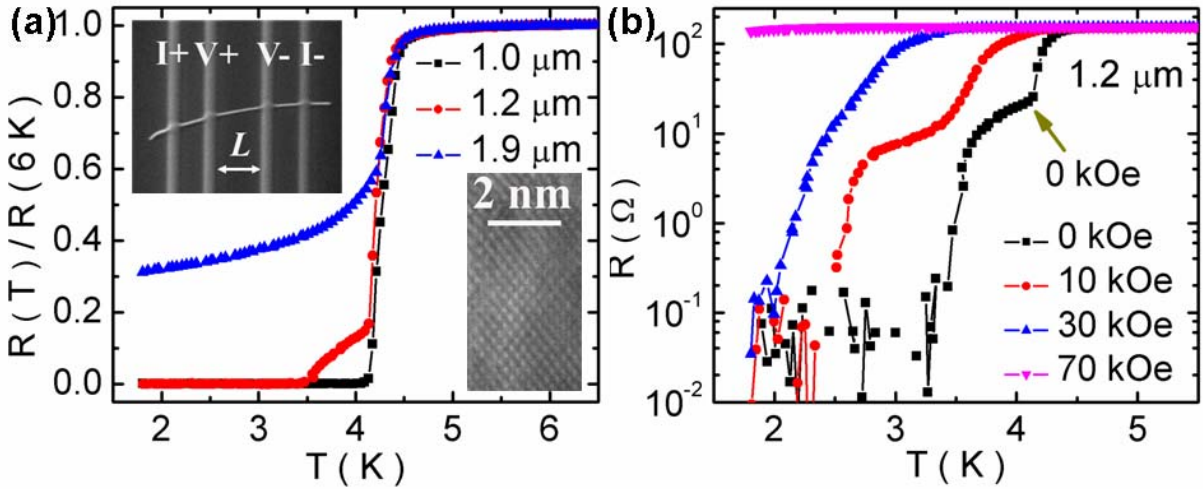

**Figure 1.** (a) Resistance as a function of temperature for individual 70 nm diameter crystalline Au nanowires with lengths (L) of 1 μm, 1.2 μm and 1.9 μm. The vertical scale is normalized to the resistance at T=6 K. The top left inset is a scanning electron micrograph (SEM) of an individual 70 nm Au nanowire contacted by four FIB-deposited superconducting W electrodes. The bottom right inset shows a high-resolution transmission electron microscope (TEM) image of a free-standing crystalline Au nanowire showing atomic structure. (b) Resistance vs. temperature plots for the 1.2 μm Au nanowire at different magnetic fields applied perpendicular to wire axis.

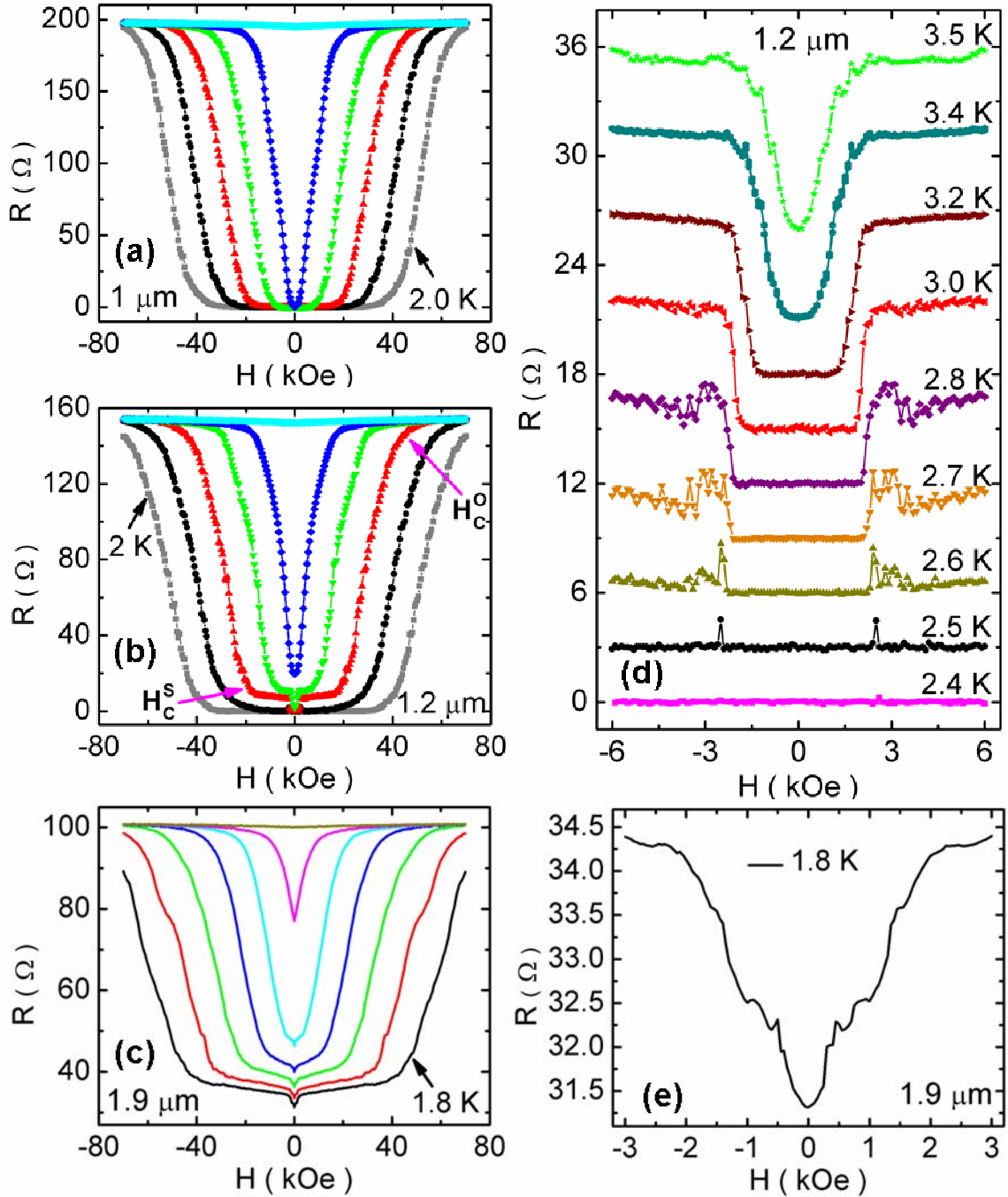

**Figure 2.** (a) Magnetoresistance of the 1 μm Au nanowire, from bottom to top, at 2.0 K (gray), 2.5 K (black), 3.0 K (red), 3.5 K (green), 4.0 K (blue), and 5.5 K (cyan). (b) Magnetoresistance of the 1.2 μm Au nanowire, from bottom to top, at 2.0 K (gray), 2.5 K (black), 3.0 K (red), 3.5 K (green), 4.0 K (blue), and 5.5 K (cyan). (c) Magnetoresistance of the 1.9 μm Au nanowire, from bottom to top, at 1.8 K (black), 2.3 K (red), 2.8 K (green), 3.3 K (blue), 3.8 K (cyan), 4.3 K (magenta), and 5.8 K (dark yellow). (d) Close-up view of (b) near zero magnetic field at different temperatures. The curves are offset for clarity; except for 3.5 K, the resistance in all plots is zero at zero field. (e) Close-up view of (c) near zero magnetic field at 1.8 K.

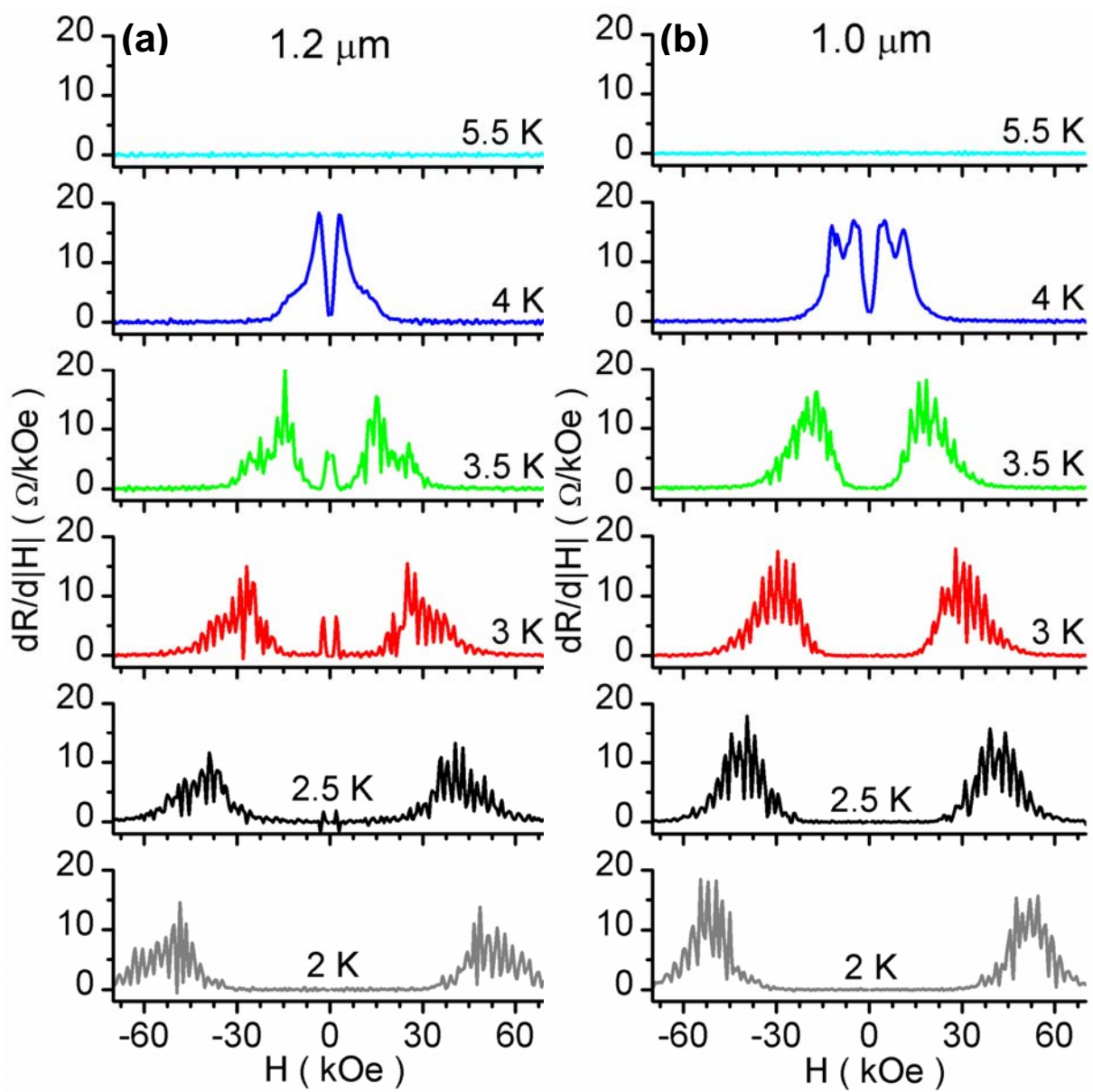

**Figure 3.** dR/d|H| of the (a) 1.2 μm and (b) 1.0 μm Au nanowires.

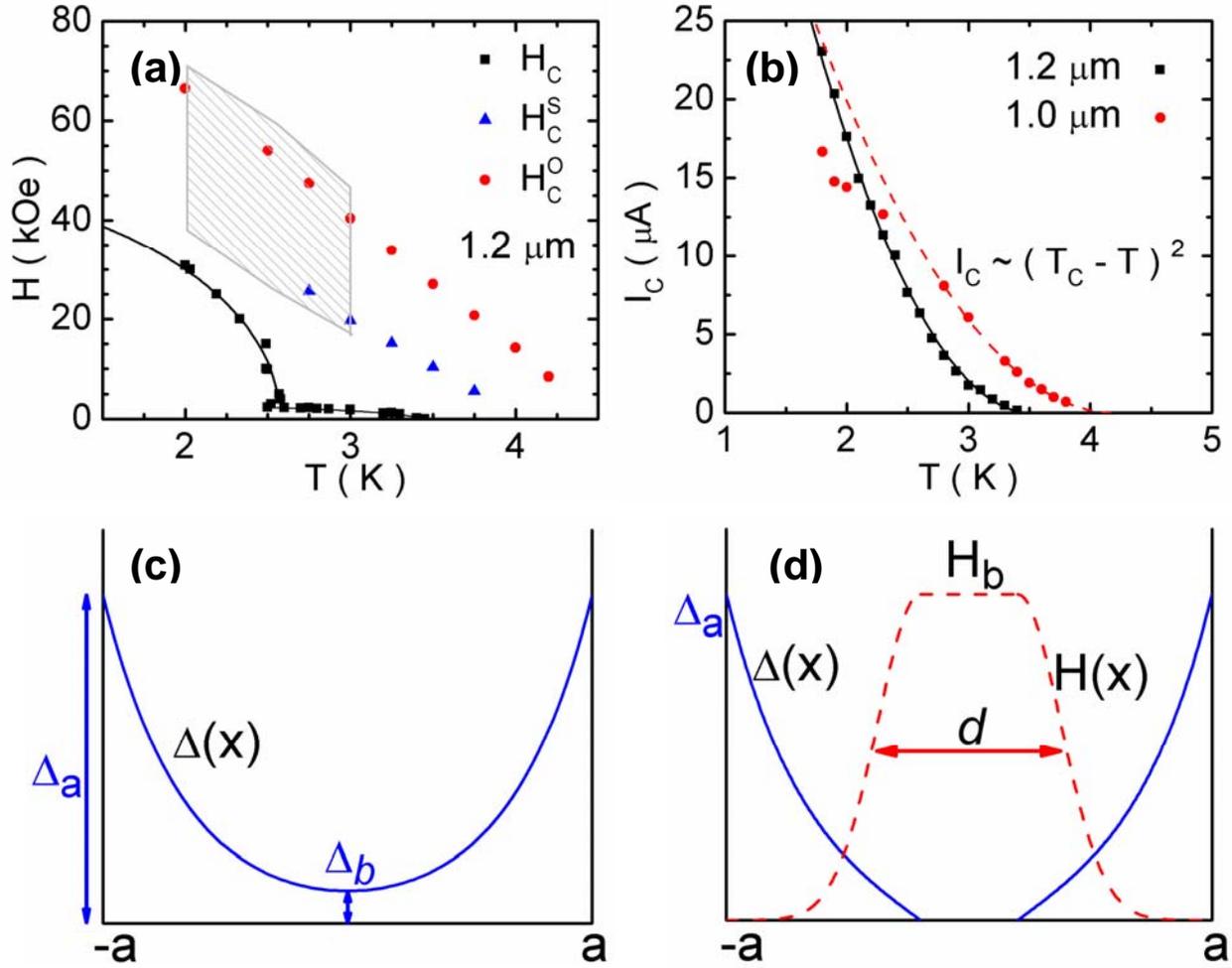

**Figure 4.** (a) The phase diagram of the 1.2 μm Au nanowire. The black squares define the phase boundary separating the zero and finite resistance states. The meaning of the blue triangles, $H_C^S$, and red circles, $H_C^O$, is described in the text. The oscillations in dR/dH occur in the shaded region. (b) Critical current $I_C$ as a function of the temperature for both 1 and 1.2 μm wires. The fitted curves follow the relation $I_C \propto (T_C - T)^2$. (c) The gap function $\Delta(x)$ of the wire as given by equation (1). (d) The gap function in the wire at the breakdown field $H_b$. $\Delta_b$ is suppressed at this transition while $\Delta_a$ essentially remains unaffected. The quantity $d$ is the length of the segment in which the magnetic field is roughly the same as the applied field.